\documentclass[preprint2]{aastex}

\newcommand{\vsini}{\mbox{$v\sin i$}}

\newcommand{\kms}{km s$^{-1}$}
\newcommand{\Teff}{$T_{\rm eff}$}
\newcommand{\logg}{$\log\thinspace g$}
\newcommand\vmac{\mbox{$v_{\rm mac}$}}
\newcommand\vmic{\mbox{$v_{\rm mic}$}}

\def\ltsima{$\; \buildrel < \over \sim \;$}
\def\simlt{\lower.5ex\hbox{\ltsima}}
\def\gtsima{$\; \buildrel > \over \sim \;$}
\def\simgt{\lower.5ex\hbox{\gtsima}}

\begin{document}

\title{The Magnetic Fields of Classical T Tauri Stars}

\author{Christopher M. Johns--Krull\altaffilmark{1}}
\affil{Department of Physics \& Astronomy, Rice University, 6100 Main St.
       MS-108, Houston, TX 77005}
\email{cmj@rice.edu}

\altaffiltext{1}{Visiting Astronomer, Infrared Telescope Facility, operated
for NASA by the University of Hawaii}

\begin{abstract} 

We report new magnetic field measurements for 14 classical T Tauri stars 
(CTTSs).  We combine these data with one previous field determination in
order to compare our observed field strengths with the field strengths
predicted by magnetospheric accretion models.
We use literature data on the stellar mass, radius, rotation period, and
disk accretion rate to predict the field strength that should be present on
each of our stars according to these magnetospheric accretion models.  We
show that our measured field values do not correlate with the field 
strengths predicted by simple magnetospheric accretion theory.  We also use
our field strength measurements and literature X-ray luminosity data to test
a recent relationship expressing X-ray luminosity as a function of surface 
magnetic flux derived from various solar feature and main sequence star 
measurements.  We find that the T Tauri stars we have observed have
weaker than expected X-ray emission by over an order of magnitude on 
average using this relationship.  We suggest the cause for this is actually
a result of the very strong fields on these stars which decreases the 
efficiency with which gas motions in the photosphere can tangle magnetic
flux tubes in the corona.

\end{abstract}


\section{Introduction} 

It is now generally accepted that accretion of circumstellar disk material
onto the surface of a classical T Tauri star (CTTS) is controlled by a
strong stellar magnetic field (e.g. see review by Bouvier et al. 2006).
It is within the disks around these low mass pre-main sequence stars that 
solar systems similar to our own form.  Understanding the processes through 
which young stars interact with and eventually disperse their disks is 
critical for understanding the rotational evolution of stars and the formation
of planets.  A key question is to understand how young stars can accrete
large amounts of disk material with high specific angular momentum, yet
maintain rotation rates that are observed to be relatively slow
(e.g. Hartmann \& Stauffer 1989, Edwards et al. 1994).  This problem is solved
in current magnetospheric accretion models by having the stellar magnetic 
field truncate the inner disk, typically near the corotation radius, and 
channel the disk material onto the stellar surface, most often at high stellar
latitude.  

Support for magnetospheric accretion in CTTSs is significant.  Current models
can account for the relatively slow rotation of most CTTSs (Camenzind 1990;
K\"onigl 1991; Collier Cameron \& Campbell 1993; Shu et al. 1994; Paatz \& 
Camenzind 1996; Long, Romanova, \&
Lovelace 2005).  Studies of the spectroscopic and photometric variability of 
CTTSs have been interpreted in terms of magnetically controlled accretion 
(e.g. Bertout, Basri, \& Bouvier 1988; Bertout et al.  1996; Herbst, 
Bailer--Jones, \& Mundt 2001; Alencar, Johns--Krull, \& Basri 2001) with the 
magnetic axis inclined to the rotation axis in some cases (e.g. Kenyon et al.
1994, Johns \& Basri 1995, Bouvier et al. 1999, Romanova et al. 2004).  
Models of high 
resolution Balmer line profiles, computed in the context of magnetospheric 
accretion, reproduce many aspects of observed line profiles (Muzerolle, 
Calvet, \& Hartmann 1998, Muzerolle et al. 2000).  Such models can also
reproduce aspects of the observed \ion{Ca}{2} infrared triplet lines
(Azevedo et al. 2006).  T Tauri stars (TTSs) are
observed to be strong X-ray sources indicating the presence of strong magnetic
fields on their surfaces (see Feigelson \& Montmerle 1999 for a review), and 
several TTSs have been observed to have strong surface magnetic fields 
(Basri, Marcy, \& Valenti 1992; Guenther et al. 1999; Johns--Krull, Valenti, 
\& Koresko 1999b, hereafter Paper I; Johns--Krull, 
Valenti, \& Saar 2004, hereafter Paper II; Yang, Johns--Krull, \& Valenti 
2005, hereafter Paper III), and strong magnetic fields have been observed in 
the formation region of the \ion{He}{1} emission line at 5876 \AA\ 
(Johns--Krull et al.  1999a; Valenti \& Johns--Krull 2004; Symington et al. 
2005), which is believed to be produced in a shock near the stellar surface 
as the disk material impacts the star (Beristain, Edwards, \& Kwan 2001).

Despite these successes, open issues remain.  Most current theoretical models
assume the stellar field is a magnetic dipole with the magnetic axis aligned
with the rotation axis.  However, recent spectropolarimetric measurements
show that the fields on TTSs are probably not dipolar, either aligned with the
rotation axis or not (Johns--Krull et al. 1999a; Valenti \& Johns--Krull 
2004; Daou, Johns--Krull, \& Valenti 2005; Yang, Johns--Krull, \& Valenti
2006).  On the other hand, it is expected
that even for complex magnetic geometries, the dipole component of the field 
should dominate at distance from the star where the interaction with the disk 
is taking place, so this may not contradict current theory.  In the case of
the Sun, the dipole component appears to become dominant at $2.5 R_\odot$ or 
closer (e.g. Luhmann et al. 1998).  For expected disk truncation radii of 
3 -- 10 $R_*$ in CTTSs (see below and Table \ref{sample}), this suggests the 
dipole component will govern the stellar interaction with the disk.  
Additionally, Gregory et al. (2006) show that accretion can occur from a 
truncated disk even when the stellar field geometry is quite complex; however,
no study has considered the torque balance between a star and its disk in the
case of a complex stellar field geometry.  

Magnetospheric accretion models for CTTSs have been developed by a number of
investigators.  K\"onigl (1991) first applied the work of Ghosh and Lamb 
(1979) to CTTSs, showing that an equilibrium state could exist if the stellar
field was strong enough.  The field would truncate the inner disk and the
star would spin at the same angular velocity as the disk where it was
truncated just inside the corotation radius.  Additional authors have
have studied this equilibrium state, utilizing different assumptions regarding
the details of how the stellar field couples to the disk.  Paper I
examined the models of K\"onigl (1991), Collier Cameron and
Campbell (1993), and Shu et al. (1994) and presented equations from each
work which predict the stellar magnetic field strength given the stellar 
mass ($M_*$), radius ($R_*$), rotation period ($P_{rot}$), and mass accretion
rate ($\dot M$).  These three studies each assume the stellar field is a dipole
aligned with the rotation axis.  In this simple case, there should be no 
photometric or spectroscopic variability associated with accretion assuming
the disk is uniform.  As mentioned above, the observations clearly show 
substantial accretion variability, and they also show that the fields are
probably not aligned dipoles, at least at the stellar surface.  Nevertheless,
it may well be that equilibrium models mentioned above give an indication of
the mean, or time averaged behavior of a CTTS.  

Several authors have performed
numerical simulations of magnetospheric accretion.  Typically, aligned stellar
dipole fields are assumed and the simulations examine the instabilities
that can result between the star, its magnetosphere, and the disk.  Many
studies find that winds and/or jets are produced (e.g. Hayashi, Shibata, \& 
Matsumoto 1996; Goodson, Winglee, \& B\"ohm 1997; Miller \& Stone 1997; 
Goodson, B\"ohm, \& Winglee 1999; Ferreira, Pelletier \& Appl 2000; Romanova
et al. 2005).  These studies usually 
examine only a few select cases and generally do not address how the rotation
of the star may evolve: it is a fixed parameter of the model.  These 
simulations may be testable through variability studies (see Goodson et al. 
1999).   Recently, Long et al. (2005) present numerical simulations, again
assuming an aligned dipole field geometry, with the aim of determining if
a time averaged equilibrium rotation rate is established.  Despite variability
in the accretion and wind flows, these authors find that the star attains
an equilibrium rotation rate with a period equal to that of the disk where it
is truncated inside, but close to, the corotation radius.  Long et al. 
(2005) use the results of their simulations to derive a relationship for
the equilibrium rotation period of CTTSs (their equation 12).  This can
be solved for the stellar magnetic field strength and results in an equation
with the same dependence on the stellar ($M_*$, $R_*$, $P_{rot}$) and
accretion parameters ($\dot M$) as found for K\"onigl (1991) and Shu et al.
(1994) as given in Paper I.  It should be noted however,
that Matt and Pudritz (2004, 2005a) have recently asserted that the spin
down torque produced on the star in magnetospheric accretion models may be
up to an order of magnitude lower than most previous investigators have
calculated.  Matt and Pudritz (2005) suggest that an accretion powered
{\it stellar} wind is responsible for spinning down CTTSs.  

Given the theoretical uncertainty involved with magnetospheric accretion, we 
can try to use observations to test whether the disk locking scenario 
proposed by the equilibrium models actually holds 
in CTTSs.  One of the first attempts at this is the work of Stassun et al. 
(1999) who find no correlation between rotation period and the presence
of an infrared (IR) excess indicative of a circumstellar disk in a sample of
254 stars in Orion.  However, IR excess alone is not a good measure of the 
accretion rate.  Muzerolle, Calvet, and Hartmann (2001) note that current 
theory predicts a correlation between rotation period and mass accretion rate
which they do not observe.  Muzerolle et al. (2001) suggest that variations in 
the stellar magnetic field strength from star to star may account for the lack
of correlation.  Indeed, Paper I emphasizes that there are
several stellar and accretion parameters that enter into the equilibrium
relationship, and the stellar magnetic field remains the quantity measured 
for the fewest number of CTTSs.  Due primarily to this lack of data,
Johns--Krull and Gafford (2002) made the assumption that the field in fact
does not vary significantly from star to star and they looked for correlations
among the remaining stellar and accretion parameters.  The predictions based
on aligned dipole field geometries are not present in the data; however, if
the dipole assumption is dropped, Johns--Krull and Gafford showed that the
predicted correlations are present in the samples they examined from CTTSs
in Taurus.

The success of the Johns--Krull and Gafford (2002) study depends on the
constancy of the magnetic field from one TTS to the next.  From a dynamo
perspective, this may be a good assumption.  For cool main sequence stars
there is a well defined positive correlation between magnetic activity indices
(e.g.  \ion{Ca}{2} H\&K emission, X-ray emission) and inverse Rossby number
(the convective turnover time divided by the rotation period) with saturation
setting at large inverse Rossby number (e.g. Vilhu 1984, Noyes et al. 1984).
Johns--Krull, Valenti, and Linsky (2000) showed that due to their long 
convective turnover times and relatively short rotation periods, TTSs generally
all lie in the saturated portion of this relationship.  On the other hand, most
TTSs are fully convective or nearly so, so a solar-like interface dynamo
(Durney \& Latour 1978) is likely very inefficient or non-existent in these
stars.  Another possibility is a convective/turbulent dynamo such as 
originally proposed by Durney, DeYoung, \& Roxburgh (1993).  Recent studies
by Chabrier and K\"uker (2006) and Dobler, Stix, and Brandenburg (2006) do 
predict that dynamo action in fully convective stars does correlate with 
rotation rate.  A final potential contributor to
the fields of TTSs is primordial fields entrained in the star during the
star formation process (e.g. Tayler 1987, Moss 2003).  However, little
is known about how these fields might vary from star to star in a given star
formation region.  As a result, additional measurements of the magnetic
field strength on a sizeable sample of CTTSs are required to test current
magnetospheric accretion models.

To that end, we have been conducting observations of TTSs at the NASA Infrared
Telescope Facility (IRTF) to measure the Zeeman broadening of several K
band \ion{Ti}{1} lines which are prominent in the spectra of cool stars.
The first use of this line to measure a field on a TTS is in Paper I, where
we lay out most of the basics of our observation and analysis technique.
Details have been expanded upon somewhat in Papers II and III.  In particular,
in Paper III, it was shown that errors in effective temperature of $\sim 200$
K as well as errors of 0.5 dex in log$g$ result in errors of less than
10\% in the derived mean magnetic field on TTSs.  This is because we are
measuring actual Zeeman broadening of the photospheric absorption lines
instead of some secondary effect such as a change in line equivalent width.
As a result, it is no longer necessary to perform the detailed analysis of
optical high resolution spectra to obtain the best estimates of \Teff\ and
\logg\ in order to obtain a good magnetic field measurement.  What is needed
is a good estimate of \vsini, and all the stars in our sample have reliable
measurements from the literature.  
Here we focus on CTTSs for which we have the ancillary data
required to predict the stellar magnetic field strength from magnetospheric
accretion models.  To do that, we use the equations (1-3) given in Paper I 
for the studies of K\"onigl (1991), Collier Cameron and Campbell (1993),
and Shu et al. (1994), and we use equation (12) of Long et al. (2005).  In
order to make the field predictions, we need stellar and accretion parameters.
Rotation periods come from Bouvier et al. (1993, 1995), with the exception
of TW Hya which comes from Mekkaden (1998).  Spectral type, stellar 
radius, and the mass accretion rate for most stars come from Gullbring et al. 
(1998).  For DG Tau, DH Tau, and T Tau we take these values from White and 
Ghez (2001).  In some cases (e.g. DH Tau), the spectral type is listed as 
spanning two subtypes (e.g. M1-2) and here we average the subtypes (e.g. M1.5).
For all stars, we use pre-main sequence evolutionary tracks of Siess, Dufour,
and Forestini (2000) to derive the stellar mass.
Our sample of stars is given in Table \ref{sample}.  Included in Table 
\ref{sample} is the observed X-ray luminosity for each star.  These come
from Neuhauser et al. (1995) except for DF Tau, DH Tau, and GM Aur which 
come from Damiani et al. (1995); and for TW Hya which comes from Kastner et
al. (2002).  The goal of this paper is to measure the mean magnetic field 
strength for all the stars in Table \ref{sample} so that the observations 
can be compared with the theoretical predictions.  Some of the stars in our
sample are known close binaries; however, in these cases one member of the
binary dominates the optical and IR light, and it is this member to which 
the various stellar, accretion, and magnetic parameters apply.
Included in our sample
is TW Hya which we do not analyze here, but was analyzed in Paper III.  In 
\S 2 we describe our observations and data reduction.  The magnetic field 
analysis is described in \S3.  In \S 4 we give our results and we compare 
our measured and the predicted field strengths.  We also compare our measured
field strengths and the observed X-ray luminosity of our targets to 
theoretical predictions based on main sequence stars.  A discussion of our 
results is given in \S 5.

\section{Observations and Data Reduction}

All spectra were obtained with the CSHELL spectrometer (Tokunaga et
al.  1990, Greene et al.\ 1993) on the 3.0 m telescope at the NASA IRTF
on Mauna Kea in Hawaii. Observations occurred during four observing runs: 
I) 10--12 December 1996, II) 14--20 December 1997, III) 7--12 January 2000,
and IV) 23--28 November 2004 (see Table \ref{obs}).  On the first run a
1\farcs0 slit gave a FWHM of $\sim 4.7$ pixels on the $256 \times 256$ InSb
array detector, corresponding to a spectral resolving power of 
$R \equiv \lambda /\Delta\lambda \sim 21,500$.  For the second
and fourth runs, a 0\farcs5 slit yielded a FWHM of $\sim 2.8$ pixels on
the detector giving $R \sim 36,000$. For the third observing run, the same 
0\farcs5 slit yielded a spectral resolving
power of only $R \sim 27,000$. [Soon after run III, CSHELL was serviced
to restore the spectral resolution.] CSHELL uses a continuously variable
filter (CVF) to isolate individual orders of the echelle grating. Each
star was observed in three settings, each covering $\sim 0.0057$
\micron. The first setting (1) contains two strong \ion{Ti}{1} lines at
2.2211 and 2.2233 \micron. The second (2) contains two more strong
\ion{Ti}{1} lines at 2.2274 and 2.2311 \micron. The third setting (3)
centered at 2.3120 \micron\ contains $\sim 9$ CO lines from the $v =
2-0$ first overtone band. 

Each star was observed at two positions along the slit separated by
10\arcsec.  Multiple pairs of exposures were made of each CTTS.  Total
exposure time varied some from object to object and wavelength setting to
wavelength setting.  The minimum exposure time was 20 minutes and the maximum
was 1.5 hours.  A typical total exposure time was 1 hour.  Taking the 
difference of each image pair with the star moved along the slit removed 
detector bias, dark current, and the average background due to night sky
emission. Difference images were flat fielded using a normalized
spectrum of an internal continuum lamp that fully illuminated the
slit.  Stellar spectra were extracted following the procedure described
in Paper I, which includes the determination of an
oversampled slit function and optimal extraction of the spectrum.
Wavelengths were determined by fitting $n\lambda$ as a function of line
position for 7--10 lamp emission lines and then dividing by the order
number, $n$, of the desired spectral region. Calibration lines from several
orders were obtained by changing the order-sorting CVF while keeping
the grating fixed.  In addition to CTTSs, each night we observed hot, rapidly
rotating stars spanning a range of airmasses.  Using these data, all spectra 
shown in this paper have been corrected for telluric absorption.
For each CTTS in our sample, Table \ref{obs} gives a log of when
the star was observed and which wavelength settings were covered.

\section{Analysis}

     The most successful approach for measuring fields on late--type 
stars in general has been to measure Zeeman broadening of spectral lines in
unpolarized light (e.g., Robinson 1980; Saar 1988; Valenti, Marcy, \& Basri
1995; Johns--Krull \& Valenti 1996; Papers I, II, \& III).  In the presence
of a magnetic field, a given spectral line will split up into a number
of components depending the atomic structure of the levels contributing to
the line.  Taking into account rotational and turbulent broadening, a
strong magnetic field typically produces a change in the shape of magnetically
sensitive line profiles.
There are several \ion{Ti}{1} lines in the K band which are excellent probes
of magnetic fields in late-type stars (e.g. Saar \& Linsky 1985), and here 
we observe 4 of them (see \S 2).  In addition to the 4 strongly Zeeman
sensitive \ion{Ti}{1} lines, our wavelength settings also cover a relatively
weak, but detectable, \ion{Sc}{1} line at 2.2265 $\mu$m which has 
reduced Zeeman sensitivity due to lower Land\'e-$g$ values.  We also
observe several magnetically insensitive CO lines near 2.313 $\mu$m which
serve as a check on our values of \vsini\ and turbulent broadening terms.
Since we are observing CTTSs, there is some concern that these lines could
form in the disk and trace disk kinematics instead of stellar properties.
This issue was studied for CO by Casali and Eiroa (1996) and Johns--Krull
and Valenti (2001).  Both studies find that the K band CO lines of the
vast majority of CTTSs arise in the stellar photosphere, and the CO lines
observed below are fully consistent with formation on the star.  Based on
their behavior as a function of effective temperature, the \ion{Ti}{1} lines
in the K band appear to trace similar temperature material as that
traced by the CO (Wallace \& Hinkle 1997).  In addition, the limited 
studies of TTS which show no K band excess (Johns--Krull et al. 2004,
Yang et al. 2005) reveal \ion{Ti}{1} and CO line profiles with the same 
qualitative behavior as that found below: CO lines well matched by 
rotationally broadened stellar photosphere models and excess
broadening in the magnetically sensitive \ion{Ti}{1} lines.  We therefore
assume that the line profiles are dominated by the stellar photosphere.

The goal in this paper is to measure the magnetic field on a sample of CTTSs
by modeling the profiles of their K band photospheric absorption 
lines.  Our spectrum synthesis code and detailed analysis technique for 
measuring magnetic fields on TTSs is described in Papers I -- III.  Here,
we simply review some of the specific details relevant to the results shown
here.  In order to synthesize the stellar spectrum, we must first specify
the atmospheric parameters: effective temperature (\Teff), gravity (\logg), 
metallicity ([M/H]), microturbulence (\vmic), macroturbulence (\vmac), and
rotational velocity (\vsini).  Following Papers I --- III,
a fixed value of 2 \kms\ is adopted for \vmac.  Valenti, Piskunov,
\& Johns--Krull 1998 found that microturbulence and macroturbulence 
were degenerate in M dwarfs, even with very high quality spectra.  Therefore,
microturbulence is neglected here,
allowing \vmac\ to be a proxy for all turbulent broadening.
Solar metallicity is assumed for all stars.
We use \vsini\ values measured using
the CO spectra presented here by Johns--Krull and Valenti (2001) for all
the target stars analyzed in that paper.  We adopt
\vsini\ values from Hartmann et al. (1986) for CY Tau and GM Aur, and we
adopt the \vsini\ value of Basri and Batalha (1990) for DG Tau.
The \Teff\ is based on spectral type using the calibration of Johnson
(1966) and \logg\ is estimated by placing objects in the HR diagram.
This typically produces uncertainties of $< 200$ K in \Teff\ and $< 0.5$
dex in \logg\ which translate into uncertaities of 10\% or less in the
derived magnetic field values (Paper III).
Once we have estimates of \Teff\ and \logg, we use the
``next generation" (NextGen) model atmosphere (Allard \& Hauschildt 1995)
from the published grid which is closest to these values.  In the range of 
interest here, the NextGen models are on a grid every 100 K (e.g. 3900, 4000,
4100) in \Teff\ and every 0.5 dex (e.g. 3.5, 4.0, 4.5) in \logg.  For BP Tau, 
we use the NextGen model with \Teff$=4100$ K which is closest in temperature 
to the value we derived in Paper I (its spectral type of K7 suggests 
\Teff$=4000$ K).

Lastly, we solve for the K band veiling in each star 
along with the stellar magnetic field properties.
We initially decided to fit for a single veiling parameter which would apply to
all three spectral regions we observe in each star.  This turned out to not
be practical, as a few stars could not be well fit by a single veiling 
value.  The different wavelength regions for our stars
were observed with a time interval between settings of several hours up to
a couple of days.  Classical TTSs regularly show significant variations in 
their K band flux on timescales as short as a day (and occasionally shorter),
likely as a result of accretion variability (Skrutskie et al. 1996, Eiroa et
al. 2002).  We therefore decided to let the veiling be a free parameter for
each star in each of the three wavelength settings analyzed here.

\begin{figure} 
\epsscale{.75}
\plotone{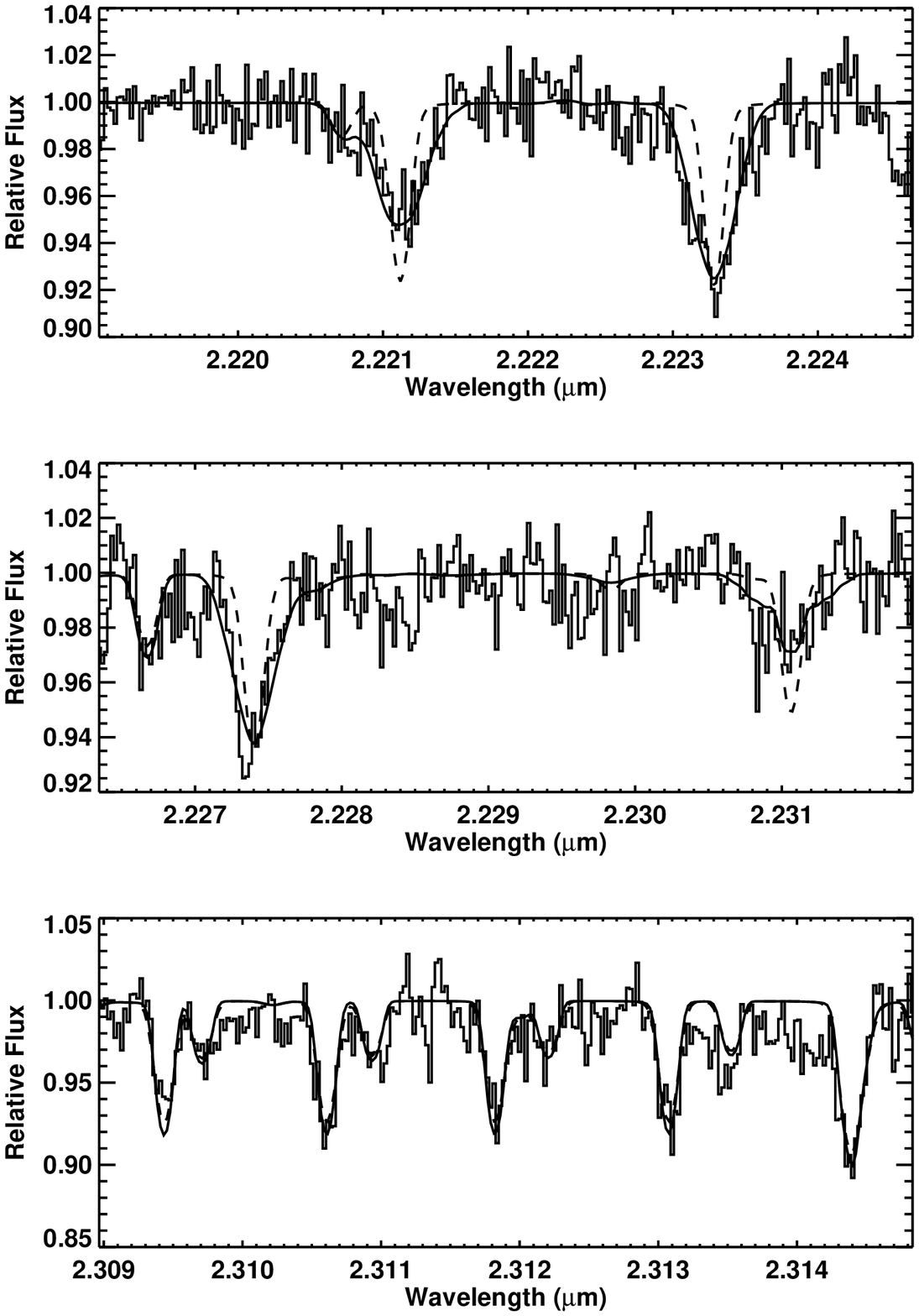}
\caption{K band spectra of DK Tau are shown in the black histogram.  The 
top 2 panels show Zeeman sensitive Ti I lines and a Sc I line.  The bottom 
panel shows Zeeman insensitive CO lines.  The dashed curve shows a model with
no magnetic field.  The smooth solid curve shows the model fit for a star 
covered by a mixture of regions which include 0, 2, 4, and 6 kG fields.  Here,
the mean field strength averaged over the stellar surface is 2.64 kG.
\label{highb}}
\end{figure}

\begin{figure} 
\plotone{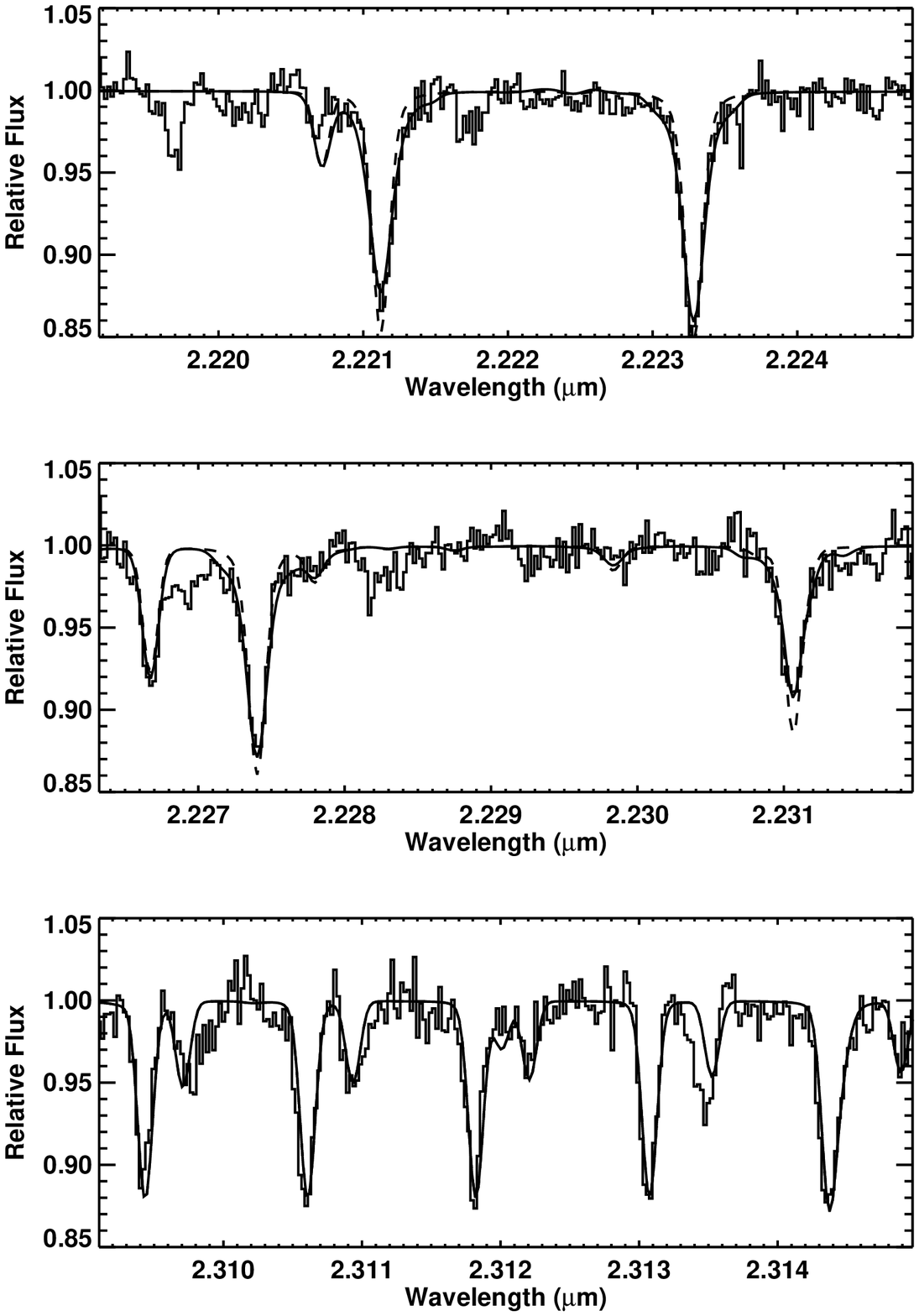}
\caption{K band spectra of DE Tau are shown in the black histogram.  The 
top 2 panels show Zeeman sensitive Ti I lines and a Sc I line.  The bottom 
panel shows Zeeman insensitive CO lines.  The dashed curve shows a model with
no magnetic field.  The smooth solid curve shows the model fit for a star 
covered by a mixture of regions which include 0, 2, 4, and 6 kG fields.  Here,
the mean field strength averaged over the stellar surface is 1.12 kG.
\label{lowb}}
\end{figure}

The free parameters of our fit to the observed spectra are then the value
of the K band veiling in each wavelength region (this is the only free
parameters in the CO wavelength region) and the magnetic field properties
of the star.  It was found in Papers I -- III that the spectra of the TTSs
studied there could not be fit with a single value of the magnetic field
strength.  Instead, a distribution of magnetic field strengths is required.
It was also found that fits to the spectra are degenerate in the derived
field values unless we limit the fit to specific values of the magnetic
field strength, separated by about 2 kG, which is the approximate ``magnetic
resolution" of our data.  Therefore, we use the same limitations when
fitting the data presented here.  We assume the star is composed of regions
of 0, 2, 4, and 6 kG magnetic field, and we solve for the filling factor
of each of these components.  The different regions are assumed to be 
well mixed over the surface of the star -- different components are not
divided up into well defined spots or other surface features.  Another key
assumption (which will be discussed further in \S 5) is that the temperature
structure in all the field regions is assumed to be identical for the
purpose of spectrum synthesis:  again, the fields are not confined to cool
spots or hot plage--like regions.  In total then, we fit for 6 free parameters:
the 3 veiling values and the filling factor of the 3 non-zero field
regions (the sum of the filling factors must be 1.0).  In Figures \ref{highb}
and \ref{lowb} we show our spectral fits to two of our CTTSs, the first
(DK Tau) with a high average surface field, and the second (DE Tau) with a
low average surface field.
We characterize the field on all our CTTSs by computing the mean
magnetic field from our fits: $\bar B = \Sigma B_i f_i$ where $B_i$ is the
value of the field (0, 2, 4, 6 kG) in each fitted region and $f_i$ is the
filling factor of these field components.
The formal uncertainty in the fit to all our stars is quite small.  The
true uncertainty is dominated by uncertainties in the field resulting from
errors in our adopted temperature and gravity, which as discussed above
are $\sim 10$\%.  We therefore adopt this value for the uncertainty in our
mean magnetic field for each star.  

While DK Tau's \ion{Ti}{1} lines show obvious broadening relative to the
nonmagnetic model in Figure \ref{highb}, the situation is a little less clear
for DE Tau in Figure \ref{lowb}.  Blowing up the plot, it is apparent that
all the \ion{Ti}{1} lines are broader than the null field model which is a
good match to the width of the CO lines.  This extra broadening in the
\ion{Ti}{1} lines can be quantified by a cross correlation analysis as well.
We construct a kernel which is the zero field model but with a \vsini\ $=1.0$
\kms\ instead of 7 \kms\ which is the measured value for DE Tau.  We then
cross-correlate this kernel with the observed DE Tau spectrum and fit the
result with a Gaussian profile to determine the FWHM of the cross correlation
function.  Doing this on the CO portion of the spectrum (lower panel of
Figure \ref{lowb}), we find a FWHM$ = 6.35$ pixels.  Cross correlating the
1.0 \kms\ kernel with the actual fit to the CO lines (dashed line of Figure
\ref{lowb} where \vsini$=7.0$ \kms), we find the FWHM$ = 6.54$ pixels (which
is slightly greater - by 0.19 pixels - than for the observed spectrum).  We
can estimate the range of FWHM we should measure by taking the fit to the
spectrum shown in the dashed line and adding random noise to it to simulate
the signal-to-noise ratio (S/N) of our observed spectrum (115 for DE Tau).
We then cross correlate this synthetic data with the same kernel above and
repeat this experiment 100 times.  Doing so, we recover the a mean FWHM of
6.54 pixels and find a standard deviation in our recovered FWHM values of
0.09 pixels.  Thus, the FWHM we recover from the observations of the 
magnetically insensitive CO lines is only 2.1$\sigma$ {\it less} than that 
recovered from the CO lines of the null field model for DE Tau.  When we 
perform the same analysis on the magnetically sensitive \ion{Ti}{1} lines,
we find that the FWHM of the cross correlation function using the observations
is 9.97 pixels, whereas the null field model gives a FWHM$ = 8.66$ pixels
with a standard deviation of 0.13 pixels.  Thus the observed magnetically
sensitive \ion{Ti}{1} lines give a width 10.1$\sigma$ {\it broader} than does
the null field model.  Using the actual magnetic fit to DE Tau and again
adding 100 different realizations of random noise to it to create synthetic
data for the cross correlation analysis yields a FWHM$ = 10.12 \pm 0.17$
pixels, which is within 0.9$\sigma$ of that recovered from the observed
spectrum.  We are therefore confident in the excess width of the \ion{Ti}{1}
lines in DE Tau ($\bar B = 1.12$ kG) and in the other ``low" field measurement
stars (see below).  The analysis of Paper III shows that even for
these low mean field values, the uncertainty due to temperature and gravity
errors again lead to a mean field uncertainty of $\sim 10$\%, which is the
field uncertainty adopted here.

\section{Results}

The results of our spectral fitting are given in Table \ref{results}.  
Reported here are the derived K band veiling, $r_K$, the surface averaged
magnetic field, $\bar B$, for each star, and a prediction for the X-ray
luminosity based on the measured field values (discussed below).  One
star in our sample, GM Aur, was observed in only one wavelength setting:
the \ion{Ti}{1} setting (2) described above.  Results for this star are
somewhat less firm that for the others.

\subsection{K Band Veiling}

We determine 3 estimates of the K band veiling ($r_K$) for each star (one 
from each wavelength region) which we can compare against each other and 
against previous veiling estimates.
In Table \ref{results} we tabulate the mean value of $r_K$ determined
from the 3 settings and give as an uncertainty estimate the standard deviation
of the 3 $r_K$ values.  These veiling estimates generally agree with previous
studies (Folha \& Emerson 1999, Johns--Krull \& Valenti 2001).
We note though that by solving for the veiling in each
wavelength region separately, our magnetic field results are free of systematic
effects caused by differences in line strength between the actual star and
our spectrum synthesis.  Such differences are adjusted for by changes in the
derived $r_K$ values.  It is the {\it shape} of the \ion{Ti}{1} lines that
constrains the magnetic field vales we derive.  In support of this idea, we
looked for a correlation between the veiling values found here and the mean
fields measured.  We also looked for a correlation between the difference in
the veiling found here compared to in Johns--Krull and Valenti (2001) and
the mean field values.  No correlation is found in either case.

For one star, T Tau, our derived $r_K$ is substantially different than found
in previous studies.  At K0, T Tau is the earliest
spectral type star in our study, and one of the earliest spectral type
CTTS known.  The veiling of Johns--Krull and Valenti (2001) is derived
based on a later spectral type template star.  Similarly,
Folha and Emerson (1999) report a veiling of $r_K = 2.5$ based on comparison
with a later type template.  The \ion{Ti}{1} and CO lines used here get much
weaker in the observed spectra of dwarfs and giants as you go from K7
to K0 (e.g. Wallace \& Hinkle 1997).  This difference in intrinsic line
strength will artificially enhance the derived veiling for a K0 star when
using a later template.  The veiling we derive here is quite low, which 
might be the result of the line strength in our synthetic spectra being
too low as a result of some error in the line data or the model 
atmosphere.  To test this, we compared our model spectra for T Tau
to observed spectra of the K0 dwarf, HR 166.  This star was observed in
setting (2) during the 4th observing run (see \S 2).  Valenti and Fischer
(2005) report \Teff$ = 5221$ K and [M/H] $ = 0.16$ for this K0V star.
We observe the lines in HR 166 to be quite weak, and the 2.2311 $\mu$m 
\ion{Ti}{1} line is strongly affected by telluric absorption, so we ignore 
it here.  We measure an equivalent width of $W_{eq} = 49.4 \pm 4.8$ m\AA\ for
the 2.2274 $\mu$m based on the 4 spectra observed of HR 166.  Our 0 kG
synthetic spectrum for T Tau gives an equivalent width of $W_{eq} = 58.0$
m\AA, in good agreement with our observed value.  The fact that the 
synthetic spectrum has a larger equivalent width means that we may
actually be over estimating the veiling, not underestimating it; however,
the correction is small ($\sim 0.2$).  Therefore, we suggest our veiling
measurement for T Tau is reliable, and that previous estimates are too
large as a result of the later type template stars used for comparison.

\subsection{Magnetic Field Properties}

\subsubsection{Comparison with Magnetospheric Accretion Theory}

Table \ref{results} gives the mean magnetic field values, $\bar B$, for each
star analyzed here.  These can be compared directly with the predicted field 
strengths given in Table \ref{sample}.  Figure \ref{magcomp} compares the 
values of $\bar B$ we measure here with the predictions of magnetospheric 
accretion theory.  As mentioned above, all 4 field predictions given in Table 
\ref{sample} are very well correlated with each other.  From the standpoint
of looking for a correlation between the observed and predicted fields,
it does not matter which set of theoretical predictions we use.  Figure
\ref{magcomp} plots the predictions based on the work of Shu et al. (1994).
As is quite evident from a visual inspection of the figure, there does not
appear to be a correlation between the observed and predicted field values.
We can quantify this by computing the linear correlation coefficient (also
called Pearson's $r$ value) and its significance (e.g. Press et al. 1986).
Computing this for the data in Figure \ref{magcomp}, we find a correlation
coefficient of $r = 0.08$ which has an associated false alarm probability
of 79\%, indicating no correlation in the data at all.  While there is no
apparent correlation in Figure \ref{magcomp}, the good news is that the
observed fields generally lie on or above the line indicating equality.
Thus, the measured fields on these CTTSs are approximately the right magnitude
for magnetospheric accretion to work.  We will return to this point further
in \S 5 below.

When examining Tables \ref{sample} and \ref{results} and Figure \ref{magcomp},
worry may be aroused by the specific choice of mass accretion rates used.  It
is well known that mass accretion rates for CTTSs are difficult to measure,
and estimates for specific stars can vary by an order of magnitude or
more (see discussion in Gullbring et al. 1998).  Fortunately, the predicted
field strengths vary only as the square root of the mass accretion rate
(Johns--Krull et al. 1999b),
making them somewhat less sensitive to the problems associated with measuring
the accretion rate.  In addition, different studies generally do find good
correlation of their measured accretion rates from star to star, even if
they differ on the overall magnitude (again, see Gullbring et al. 1998).
Therefore, we have some confidence that the variation of the predicted
fields from star to star can be reasonably well estimated.  In an effort to
see how the results presented here depend on the stellar and accretion
parameters used for the stars studied, Table \ref{results2} gives the
mass accretion rates and predicted fields for 5 relatively large studies
which contain many of the stars observed here.  Since the derived mass
accretion rates depend to some extent on the mass and radius assigned to
the star, in Table \ref{results2} the predicted fields are calculated
using the mass and radius adopted for each star in the relevant study.
The studies presented in the table are Valenti et al. (1993, labelled VBJ),
Hartigan et al. (1995, labelled HEG), Gullbring et al. (1998, labelled
GHBC), Calvet and Gullbring (1998, labelled CG), and White and Ghez
(2001, labelled WG).  While accretion rates can vary by over an order of
magnitude bewteen these different studies, there is generally good correlation
in the predicted fields based on the different studies, particularly when
allowing removal of a single obviously discrepant star.  For example, 
comparing field predictions from HEG with those of GHBC produces a 
correlation ($r = 0.73$) with a false alarm probability of 1\% even though
the accretion rates differ by an order of magnitude in many cases.  Comparing
field predictions from VBJ with GHBC gives a correlation ($r = 0.74$) with
a false alarm probability of 2\% (due to the smaller number of stars in
common), and when removing the obviously discrepant GI Tau, the correlation
improves to $r = 0.99$ with a false alarm probability of $2.7 \times 10^{-6}$.

On the other hand, none
of these studies produced a good correlation with the observed mean field
strengths.  The bottom line of the table gives the correlation
coefficient and associated false alarm probability when comparing the
predicted fields from each study with the observed fields.
Plots of the observed versus predicted fields appear similar to
Figure \ref{magcomp}, showing a flat relation with no apparent correlation
even when the most discrepant point is ignored.
While, mass accretion rate is the parameter that varies the
most for each star from one study to the next, the predicted fields turn
out to be very sensitive to the adopted value of the stellar radius.  The
predicted field strengths vary as $R_*^3$ (Johns--Krull et al. 1999b),
which is usually determined by measuring the stellar luminosity and
effective temperature.  Effective temperatures usually come from spectral
type and an adopted spectral type -- effective temperature calibration;
however, this has a number of its own problems (e.g. Padgett 1996,
Huerta 2007) and is deserving of more attention.

\subsubsection{Relationship to X-ray Emission}

Stellar magnetic fields are also believed to give rise to ``activity" in
late-type stars.  Activity is typically traced by line emission or broad
band emission at high energy wavelength such as X-rays.  Pevtsov et al.
(2003) study the relationship between magnetic flux and X-ray luminosity
on the Sun and cool stars, generally finding excellent correlation between
these two quantities ranging over almost 11 orders of magnitude in each
quantity.  Pevtsov et al. (2003) include preliminary measurements for
6 of the CTTSs shown here taken from Johns--Krull and Valenti (2000) and
Johns--Krull et al. (2001).  We use the X-ray luminosity--magnetic flux
relationship of Pevtsov et al. (2003) with our magnetic field measurements
and the stellar radii given in Table \ref{sample} to calculate the expected
X-ray luminosity for each of our stars.  These predicted X-ray values are
given in the last column of Table \ref{results}.  Figure \ref{xcomp} plots
the measured X-ray luminosities from Table \ref{sample} versus these
predicted X-ray values.  All but one of the CTTSs in our sample show lower
X-ray emission than would be predicted by their magnetic field properties,
and many of them are low by more than an order of magnitude.  Only TW Hya
shows a higher level of measured X-ray emission relative to the prediction,
and perhaps significantly, this CTTSs is the star closest to the main
sequence in our sample due to its age of $\sim 10$ Myr.  We discuss this 
issue further in the next section.

\subsubsection{Field Measurements and Pressure Equipartition}

From the perspective of cool star research, the high apparent field
strengths on these CTTSs is a little surprising.  Spruit \& Zweibel (1979)
computed flux tube equilibrium models, showing that magnetic field
strength is expected to scale with gas pressure in the surrounding
non-magnetic photosphere.  Similar results were found by Rajaguru, Kurucz,
and Hasan (2002) who find that convective collapse of thin magnetic flux
tubes will set a field strength limit of $\sim 750$ G in the photosphere of
a star with \Teff$ = 4000$ K and \logg$ = 3.5$.  Convective collapse is a 
mechanism for forming partially evacuated flux tubes, and Rajaguru et al. 
(2002) find evidence that this mechanism may not operate effectively 
at low \Teff; however, they find a maximum field strength of $\sim 1300$ G if
the field is to remain in pressure equilibrium with the surrounding 
unmagnetized photosphere.   Field strengths set by such pressure equipartition 
considerations appear to be observed in G and K dwarfs (e.g. Saar 1990, 1994, 
1996) and possibly in M dwarfs (e.g. Johns--Krull \& Valenti 1996).  

TTSs have relatively low surface gravities and hence low photospheric gas 
pressures, so that equipartition flux tubes would have relatively low 
magnetic field strengths compared to cool dwarfs.  Indeed, Safier (1999) 
examined in some detail the ability of TTS photospheres to confine magnetic
flux tubes via pressure balance with the surrounding quiet photosphere, 
concluding that the maximum field strength allowable on TTSs is substantially
below the few detections reported at that time.  The maximum field strength
allowed for a confined magnetic flux tube is $B_{eq} = (8 \pi P_g)^{1/2}$
where $P_g$ is the gas pressure at the observed level in the stellar
atmosphere.  Here, we take as a lower limit (upper limit in pressure) the
level in the atmosphere where the local temperature is equal to the effective
temperature.  This is the approximate level at which the continuum forms,
with the \ion{Ti}{1} lines forming over a range of atmospheric layers
above this level.  Our computed values of $B_{eq}$ are given in Table 
\ref{results}.  In Figure \ref{eqcomp} we plot our observed mean magnetic
field strengths versus these $B_{eq}$ values for our entire sample.  
In all cases, the observed mean magnetic field is greater than the field
predicted by pressure equipartition arguments.  In many cases the observed
mean field exceeds $B_{eq}$ by a factor of more than 2.

\subsubsection{Origins of the Strong Fields}

Lastly, we compare our mean field measurements with various stellar 
parameters in an effort to uncover the origin of the strong magnetic 
fields observed on these stars.  We compare both the mean magnetic
field strength, $\bar B$, and the total magnetic flux, $\Phi$, to a number
of stellar parameters, including those thought to be important in the
dynamo generation of magnetic fields.  The results of these comparisons are
summarized in Table \ref{stelpar}.  Given in this table is the Pearson linear
correlation coefficient, $r$, and its false alarm probability, $f_p$,
which measures the degree to which the quantities being compared are
correlated with one another.  As Table \ref{stelpar} shows, there is no
significant correlation of $\bar B$ or $\Phi$ with any stellar parameter
or common dynamo variable, including the convective turnover time, $\tau_c$
[values of $\tau_c$ were kindly provided by Y.-C. Kim and were computed
using models described in Kim and Demarque (1996) with input physics 
described in Preibisch et al. (2005)].
The most significant correlation found is $\Phi$ versus $L_*$; however,
we must be careful with such a correlation due to the common role the
stellar radius plays in both variables.  The magnetic flux is $\Phi = \bar B 
R_*^2$, and the stellar radius is calculated from the measured $L_*$ and
$T_{eff}$.  Thus, we expect to find a correlation, and when comparing
$R_*$ versus $L_*$ we find $r = 0.73$ and $f_p = 0.002$.  This correlation
is stronger than that found using $\Phi$, suggesting the correlation of
$\Phi$ with $L_*$ is driven entirely by the stellar radius, with the
magnetic field playing no role.  Interestingly, the next most significant
correlation is $\Phi$ versus $\tau_c$; however, this correlation is again 
likely driven by the correlation of $R_*$ with $\tau_c$ ($r = -0.63, f_p = 
0.01$), with the magnetic field playing no role.  Within this sample of
15 CTTSs, we then find no correlation of the measured magnetic field 
values with any stellar or dynamo parameters.

\section{Discussion}

We provide new measurements of the mean magnetic field strength on 14
CTTSs.  The detected fields range from 1.12 to 2.90 kG.  These measurements
combined with previous reports of magnetic fields on TTSs (Basri et al.
1992; Guenther et al. 1999; Johns--Krull et al. 1999b, 2004; Yang et al.
2005) suggest that the strong majority of TTSs are covered by kilogauss
magnetic fields.  As found in Papers I - III, we use a distribution of magnetic
field strengths to fit the observed K band \ion{Ti}{1} profiles.  We
model this distribution as if it were a spatial distribution on the stellar
surface: the model assumes distinct regions with different field 
strengths.  A concern is that the distribution is in fact a distribution
with depth in the magnetic regions.  If this were the case, we would be
overestimating the total filling factor of magnetic regions as well as the
mean field strength.  Observations of sunspots in the K band show that the
\ion{Ti}{1} lines in this spectral region display profiles characteristic of
a single magnetic field value (e.g. Wallace \& Livingston 1992), despite the 
likely variation of the field with depth in a sunspot (e.g. Mathew et al.
2003).  Therefore, we believe the field distributions found in Papers I - III
and used here represent actual spatial distributions of field strengths on 
the stellar surface.

\begin{figure} 
\plotone{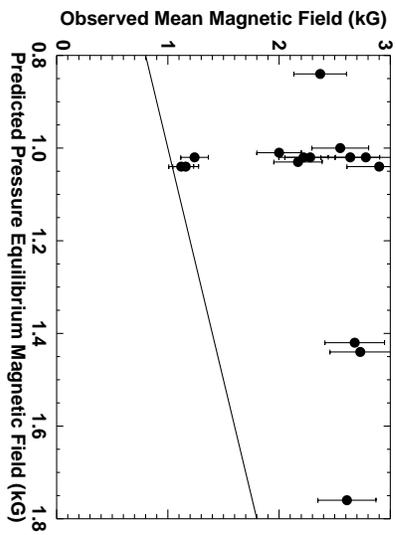}
\caption{The observed mean magnetic field on our sample of CTTSs plotted
versus the maximum allowed magnetic field strength predicted assuming 
pressure equilibrium with a surrounding non-magnetic atmosphere.  The observed
fields tend to be much stronger than the pressure equilibrium predictions,
implying magnetic filling factors near unity.
\label{eqcomp}}
\end{figure}

The observed mean field values for the CTTSs are all larger than the field
strength for these stars predicted by pressure equilibrium arguments (Figure
\ref{eqcomp}).  This suggests that the magnetic pressure dominates the gas
pressure, in some cases by a significant amount, in the photospheres of these
stars.  If this is the case, there should be no field free regions on these
stars, though we do use a 0 kG field component when fitting their spectra.
This issue is discussed in \S 4.2.3 and in Papers I and II.
In these papers, it was found that using fits with 0, 2, 4, and 6 kG
components gave essentially identical mean field measurements as those
from fits using 1, 3, 5, and 7 kG components.  Thus, the data do not demand
a field free region, and such a region may not exist on these stars.
With strong magnetic field essentially covering the entire surface of
TTSs, we expect to see some effect of the fields on the atmosphere, perhaps
analogous to the fields of sunspots inhibiting convection (e.g. Foukal 1990).
Such an effect might be diagnosed through the study of line bisectors
(e.g. Gray 1982); however, little work in this area has been done for
late K spectral types, so the intrinsic bisector shape for stars of this
spectral type is unknown.  

\begin{figure} 
\plotone{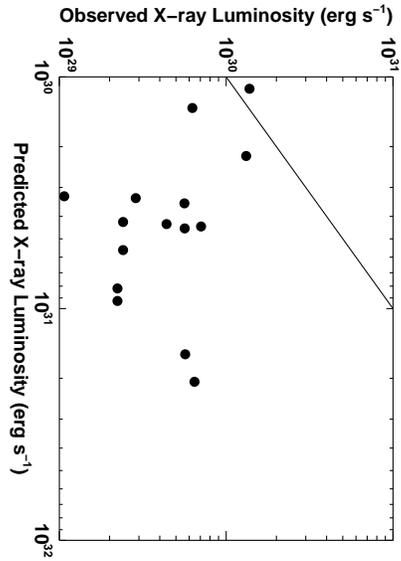}
\caption{The measured X-ray luminosity plotted against the predicted
X-ray luminosity for our sample of stars.  The line of equality is shown
by the dashed line.  The majority of these CTTSs appear underluminous in
X-rays given their magnetic field properties.
\label{xcomp}}
\end{figure}

On the other hand, we may have evidence for the
feedback of these strong fields on the star through the X-ray activity
displayed by TTSs.  Feigelson et al. (2003),  find that X-ray emission in
TTSs saturates on average at a level approximately an order of magnitude
below that displayed by main sequence stars.  In addition to this apparent
global reduction in X-ray emission of TTSs, observations also reveal that
accreting young stars as a group show X-ray emission levels that are reduced
by a factor of 2--3 relative to non-accreting young stars, even after 
dependencies on mass and age are taken into account (Flaccomio et al.
2003, Stassun et al. 2004, Preibisch et al. 2005, Feigelson et al. 2007).  
At least two explanations have been advanced that can explain this difference
between stars with and without accretion signatures.  Preibisch et al.
(2005) suggest that the higher density of mass loaded accreting field
lines prevents these portions of CTTSs magnetospheres from heating up to
temperatures required for X-ray emission.  Jardine et al. (2006) present a
model of TTS cornonae with realistic levels of complexity and show that 
this can produce a similar range of X-ray emission levels as that which
is observed in for example, the Chandra Orion Ultradeep Project (COUP). 
Jardine et al. (2006) suggest that the presence of a disk,
particularly in lower mass CTTSs, acts to strip the outer parts of the stellar
corona, thereby reducing the total X-ray emission.  This naturally predicts
an anticorrelation between X-ray emission and the presence of a disk,
at least in the lower mass young stars such as those studied here.  While 
there is no observed correlation of X-ray emission with the presence or 
absence of an inner disk as diagnosed by K-band emission for the COUP
dataset (e.g. Feigelson et
al.  2007), this coronal stripping may be related to the action of accretion.
While these ideas may explain the factor of 2--3 difference in accreting 
versus non-accreting young stars, they do not explain the order of magnitude
reduction in X-ray emission for TTSs as a whole noted by Feigelson et al. 
(2003).  Here, we find that the observed
X-ray emission from our sample of CTTSs is substantially below what is 
expected given their strong magnetic field values (Figure \ref{xcomp}),
often by more than an order of magnitude.
While the origin of coronal heating in the Sun and solar like stars is still
debated, most theories assume it is due in part to convective motions 
moving and buffeting photospheric magnetic flux tubes (e.g. Fisher et al. 
1998, Foukal 1990).  We suggest that the strong magnetic fields which appear
to cover the surface of TTSs reduce the efficiency with which convective 
motions can stress these stellar magnetic fields, which in turn results in 
reduced coronal heating relative to what would otherwise be expected.

\begin{figure} 
\plotone{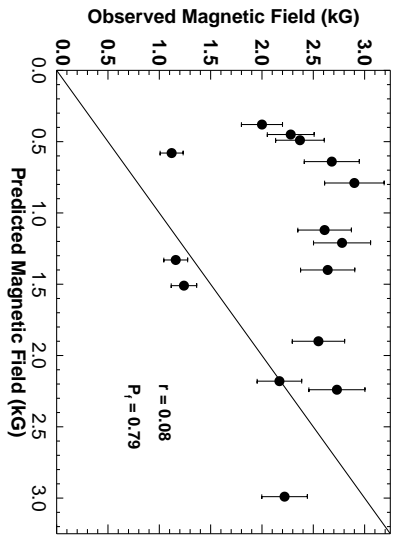}
\caption{The observed mean magnetic field strength, $\bar B$, plotted
versus the predicted equatorial magnetic field strength from the Shu et al.
(1994) treatment of the magnetospheric accretion model.
\label{magcomp}}
\end{figure}

The original motivation for this study was to test magnetospheric accretion
models for CTTSs.  Figure \ref{magcomp} shows the relationship between 
magnetic field strengths predicted from these models compared with the measured
field strengths.  Clearly, there is no correlation.
Magnetospheric accretion models typically assume an aligned dipolar magnetic
field at the surface of a CTTS.  It is now fairly clear that the fields
of TTSs are not dipolar (Johns--Krull et al. 1999a, Valenti \& Johns--Krull
2004, Daou et al. 2006, Yang et al. 2006).  The magnetic geometry must be
complicated at the surface of TTSs, as on the Sun and all other cool stars. 
However, the dipolar component may dominate at the inner edge of the 
accretion disk, several stellar radii from the surface.  This notion is
supported by observations of circular polarization in the narrow component
of the \ion{He}{1} emission line which forms near the base of the accretion
flows where the material strikes the stellar surface (Johns--Krull et al.
1999a).  Time series polarization in this line shows smooth variations 
characterized by few if any polarity reversals, and are consistent with the
idea that the disk is interacting with a dipole-like field geometry
(Valenti \& Johns--Krull 2004, Symington et al. 2005, Yang et al. 2007).

Magnetospheric accretion models use an assumed dipolar field geometry
to predict which stellar magnetic field lines couple the stellar
surface to the disk. Since the actual magnetic geometry is more complicated,
the fraction of the stellar surface with field lines coupled to
the disk (including those along which accretion will occur) can differ 
significantly from predicted values (see Gregory et al. 2006) for a 
discussion of this point.  This means the relationships given in
Paper I and used to calculate the predicted fields given in Table \ref{sample}
are actually underconstrained, even though we have now measured $\bar B$
at the surface.
However, Johns--Krull and Gafford (2002) point out that a more general
relationship for the Shu et al. (1994) model can be
determined which does not rely on the assumption of a dipolar field
geometry.  This formulation characterizes the magnetic coupling in 
terms of the surface
flux that threads the disk outside corotation, rather than the
equatorial field strength of an assumed dipole.  Valenti and Johns--Krull
(2004) use the results of Johns--Krull and Gafford (2002) to derive an 
equation for the filling factor of accretion zones, $f_a$, on CTTSs [their
equation (1)].  We use our magnetic field results with this equation to
predict $f_a$ for each of our stars.  These accretion filling factors range
from 0.57\% to 3.09\%, in good agreement with the range of observed $f_a$
values (Valenti, Johns, \& Basri 1993; Calvet \& Gullbring 1998).
The predicted $f_a$ values range over a factor of 5.4 while the observed
mean field strengths range over a factor of 2.6.  In the original 
magnetospheric accretion models, $f_a$ is essentially assumed constant and 
the predicted field strengths range over a factor of 8 (Table \ref{sample}).
We suggest that $f_a$ is more likely than $\bar B$ to vary
by a large factor.  Our measurements of $\bar B$ support this conclusion.  
Additionally, one could look for a correlation between the fields predicted
in Table \ref{sample} and the fields anchored in the accretion flows and
diagnosed by the \ion{He}{1} line.  Johns--Krull et al. (2001) do this,
finding a good correlation; however, we caution that they only had a 
sample of 4 stars.

Given that the magnetic fields on the surface of TTSs may not be
dipolar, one question that comes to mind is whether the observed fields on
CTTSs are indeed strong enough to enforce disk locking in these young stars.
The magnetospheric accretion models require some (though admittedly uncertain)
field strength in the disk, several stellar radii above the star, to enforce
disk locking.  The dipole assumption can then be used to predict what field
strength is implied at the stellar surface.  If the field geometry is much
more complex than this, a larger field strength will be required since the
higher order field components have a stronger dependence on radius than 
a dipole does.  The contribution of the various field components at the
inner disk edge is uncertain, but as mentioned above the \ion{He}{1}
observations suggest the dipole component may dominate.  This field
component is uncertain observationally, but it can be constrained through
spectropolarimetry.   Smirnov et al. (2003) report a marginal detection of
a net field of 150 G on T Tau which was not confirmed by Smirnov et al.
(2004) or Daou et al. (2006).  Yang et al. (2007) detect a net field of 
149 G on TW Hya on one night of their 6 night monitoring campaign on this
star, finding only upper limits of $\sim 100$ G on the other nights.  Other
studies also only find upper limits (3$\sigma$) of 100--200 G 
(Johns--Krull et al. 1999a, 
Valenti \& Johns--Krull 2004).  The relationship between the net field
observed spectropolarimetrically and the dipole component of the star depends
on knowing the angle between the line of sight and the dipole axis; however,
the general lack of detection suggests the dipole component on TTSs is small.
For example, a dipole with a 3 kG polar field strength (as suggested for
BP Tau by Symington et al. 2005) observed at an
angle of 45$^\circ$ would produce a net field of 690 G, well above current
detection limits.  Accretion hot spots are expected to have essentially
no effect on net field measurements from spectropolarimetry (Smirnov et 
al. 2005) since the acretion filling factors are typically $\sim 1$\%
(Valenti et al. 1993, Calvet \& Gullbring 1998). 

Compared with at least some of the predicted values in Table \ref{sample}, the
general lack of strong polarimetric field detections in TTSs might suggest the
fields are actually too weak to enforce disk locking in these stars.  However,
 the general agreement 
in magnitude between the predicted and observed accretion filling factors 
described above and in Valenti and Johns--Krull (2004) suggests that the 
trapped flux model of Shu et al. (1994, see also Ostriker \& Shu 1995) can 
still enforce disk locking with the field strengths observed here.  Indeed,
Johns--Krull and Gafford (2002) find that this trapped flux model predicts
correlations among stellar and accretion parameters that are actually
observed in the data, whereas correlations based on assuming a dipole field
geometry are not observed in the data.  As sectropolarimetric field
measurements get more precise, and the dipole component gets more constrained,
comparisons with all the field predictions in Table \ref{sample} will
become more critical.  In addition, more theoretical work is 
called for to explore whether realistic magnetic field geometries and 
strengths can produce disk locking in magnetospheric accretion models.
First steps in exploring the role of realistic field geometries in 
CTTSs accretion have been made by Gregory et al. (2006); however,
torque balance calculations still need to be made for these types of
models.

Lastly, we consider the origin of the strong fields observed in TTSs.  
Chabrier and K\"uker (2006) assert that primordial, or fossil, magnetic 
fields in fully convective stars can only survive for a timescale of
$\tau_d \sim R_*^2/\eta$, where $R_*$ is the stellar radius and $\eta$ is
the turbulent magnetic diffusivity.  The value of $\tau_d$ works out to
be on the order of 10 -- 100 years.  Thus, Charbier and K\"uker (2006)
assert that the fields of fully convective stars, such as most TTSs, 
must be produced by a turbulent dynamo.  On the other hand, Tayler (1987)
and Moss (2003) suggest that primordial fields can survive much longer
in pre-main sequence stars.  Chabrier and K\"uker (2006) and Dobler et al.
(2006) both find that dynamo action in fully convective stars should
correlate with rotation rate.  Motivated by this, we looked for correlations
between the observed magnetic fields and various stellar parameters 
including rotation (see \S 4.2.4).  No correlations were found, and 
indeed stars like CY Tau and DH Tau show very similar stellar properties
(\Teff, mass, radius, rotation rate, convective turnover time) but have
mean fields that differ by over a factor of 2.3.  In addition, the
work of Chabrier and K\"uker (2006) and Dobler et al. (2006) generally
find magnetic field strengths somewhat lower than the equipartition
values, while the observed fields are significantly stronger than this.
Additional constraints on fully convective dynamo models can also be
made by comparing predicted field geometries with those derived from
Zeeman Doppler imaging of fully convective stars.  Donati et al.
(2006) do this for an M dwarf, finding inconsistencies in the observed
field geometry relative to the predictions.  Therefore, 
it appears difficult for dynamo theories to produce the fields observed
on TTSs.  As a result, we suggest that the fields on these young stars
may indeed be primordial in origin, as suggested by Tayler (1987) and
Moss (2003).

\acknowledgements
I am pleased to acknowledge numerous, stimulating discussions with J. Valenti
on all aspects of the work reported here.  I wish to thank J. Valenti, S.
Saar, and
H. Yang for their assistance in obtaining some of the spectra used in this 
study.  I also wish to thank staff of the NASA IRTF for their hospitality and
help during my observing runs there.  Finally, I wish to acknowledge partial 
support from the NASA Origins of Solar Systems program through grant numbers
NAG5-13103 and NNG06GD85G made to Rice University.  Finally, I wish to
acknowledge many useful comments from an anonymous referee.

\clearpage
 
\begin{deluxetable}{lcccccccccc}
\tablewidth{17.5truecm}   
\tablecaption{Sample of Stars\label{sample}}
\tablehead{
   \colhead{}&
   \colhead{Spec.}&
   \colhead{$M_*$}&
   \colhead{$R_*$}&
   \colhead{$\dot{M} \times 10^8$}&
   \colhead{$P_{rot}$}&
   \colhead{$B_{eq}$}&
   \colhead{$B_{eq}$}&
   \colhead{$B_{eq}$}&
   \colhead{$B_{eq}$}&
   \colhead{$L_X$}\\[0.2ex]
   \colhead{Star}&
   \colhead{Type}&
   \colhead{(M$_\odot$)}&
   \colhead{(R$_\odot$)}&
   \colhead{$(M_\odot {\rm yr}^{-1})$}&
   \colhead{(days)}&
   \colhead{(G)}&
   \colhead{(G)}&
   \colhead{(G)}&
   \colhead{(G)}&
   \colhead{($10^{30}$ erg s$^{-1}$)}
}
\startdata
AA Tau  &K7& 0.70 & 1.74 & 0.33 & 8.20 & 1020 & 290 & 1210 & 380 &  0.439 \\
BP Tau  &K7& 0.70 & 1.99 & 2.88 & 7.60 & 1850 & 620 & 2180 & 880 &  0.707 \\
CY Tau  &M1& 0.40 & 1.63 & 0.75 & 7.90 & 1130 & 380 & 1330 & 530 & $<0.627$ \\
DE Tau  &M2& 0.32 & 2.45 & 2.64 & 7.60 &  490 & 190 &  580 & 230 & $<0.287$ \\
DF Tau  &M1& 0.40 & 3.37 & 17.7 & 8.50 &  680 & 290 &  790 & 320 &  0.646 \\
DG Tau  &K7.5& 0.65 & 2.05 & 4.57 & 6.30 & 1610 & 560 & 1900 & 760 & $<0.223$ \\
DH Tau  &M1.5& 0.36 & 1.39 & 0.10 & 7.20 &  550 & 160 &  640 & 260 &  1.318 \\
DK Tau  &K7& 0.68 & 2.49 & 3.79 & 8.40 & 1190 & 410 & 1400 & 560 & $<0.223$ \\
DN Tau  &M0& 0.51 & 2.09 & 0.35 & 6.00 &  320 & 100 &  380 & 150 &  0.564 \\
GG Tau A&K7& 0.68 & 2.31 & 1.75 & 10.3 & 1280 & 420 & 1510 & 610 & $<0.107$ \\
GI Tau  &K6& 0.93 & 1.74 & 0.96 & 7.20 & 1900 & 560 & 2240 & 900 &  0.241 \\
GK Tau  &K7& 0.69 & 2.15 & 0.64 & 4.65 &  390 & 110 &  450 & 180 &  0.241 \\
GM Aur  &K7& 0.70 & 1.78 & 0.96 & 12.0 & 2540 & 800 & 2990 &1200 &  0.562 \\
 T Tau  &K0& 2.30 & 3.31 & 4.40 & 2.80 &  420 & 110 &  490 & 200 &  0.568 \\
TW Hya  &K7& 0.74 & 1.00 & 0.20 & 2.20 &  950 & 250 & 1120 & 450 &  1.380 \\
\enddata
\tablecomments{All predicted fields are the equatorial magnetic field
strength for an assumed dipolar magnetic field.  In order, the predicted
fields come from K\"onigl (1991), Collier Cameron and Campbell (1993),
Shu et al. (1994), and Long et al. (2005).}
\end{deluxetable}

\clearpage

\begin{deluxetable}{lcc}
\tablewidth{5.5truecm}   
\tablecaption{Observing Runs\label{obs}}
\tablehead{
   \colhead{}&
   \colhead{Obs.}&
   \colhead{Wavl.}\\[0.2ex]
   \colhead{Star}&
   \colhead{Run}&
   \colhead{Set}
}
\startdata
AA Tau   &  III & 1,2,3 \\
BP Tau   &  II  & 1,2,3 \\
CY Tau   &  IV  & 1,2,3 \\
DE Tau   &  III & 1,2,3 \\
DF Tau   &  II  & 1,2,3 \\
DG Tau   &  III & 1,2,3 \\
DH Tau   &  III & 1,2,3 \\
DK Tau   &  II  & 1,2,3 \\
DN Tau   &  III & 1,2,3 \\
GG Tau A &  III & 1,2,3 \\
GI Tau   &  III & 1,2,3 \\
GK Tau   &  III & 1,2,3 \\
GM Aur   &  I   & 2 \\
T Tau    &  II  & 1,2,3 \\
\enddata
\end{deluxetable}

\clearpage

\begin{deluxetable}{lcccc}
\tablewidth{10.0truecm}   
\tablecaption{Results\label{results}}
\tablehead{
   \colhead{}&
   \colhead{}&
   \colhead{$\bar B$}&
   \colhead{$B_{eq}$}&
   \colhead{$L_X$}\\[0.2ex]
   \colhead{Star}&
   \colhead{$r_K$}&
   \colhead{(kG)}&
   \colhead{(kG)}&
   \colhead{($10^{30}$ erg s$^{-1}$)}
}
\startdata
AA Tau  & $0.52 \pm 0.09$ & 2.78 & 1.02 & 4.32 \\
BP Tau  & $1.08 \pm 0.25$ & 2.17 & 1.03 & 4.42 \\
CY Tau  & $0.65 \pm 0.21$ & 1.16 & 1.04 & 1.36 \\
DE Tau  & $1.16 \pm 0.17$ & 1.12 & 1.04 & 3.34 \\
DF Tau  & $1.37 \pm 0.15$ & 2.90 & 1.04 & 20.7 \\
DG Tau  & $2.67 \pm 0.41$ & 2.55 & 1.00 & 8.19 \\
DH Tau  & $1.77 \pm 0.49$ & 2.68 & 1.42 & 2.20 \\
DK Tau  & $1.46 \pm 0.23$ & 2.64 & 1.02 & 9.27 \\
DN Tau  & $0.54 \pm 0.09$ & 2.00 & 1.01 & 4.51 \\
GG Tau A& $0.35 \pm 0.19$ & 1.24 & 1.02 & 3.28 \\
GI Tau  & $1.00 \pm 0.11$ & 2.73 & 1.44 & 4.23 \\
GK Tau  & $1.47 \pm 0.31$ & 2.28 & 1.02 & 5.59 \\
GM Aur  & $1.25$          & 2.22 & 1.02 & 3.51 \\
 T Tau  & $0.45 \pm 0.77$ & 2.37 & 0.84 & 15.8 \\
TW Hya\tablenotemark{a}  & $0.07 \pm 0.04$ & 2.61 & 1.76 & 1.13 \\
\enddata
\tablenotetext{a}{Veiling and field measurements taken from Yang et al.
(2005).}
\end{deluxetable}

\clearpage

\begin{deluxetable}{lcccccccccccccc}
\tabletypesize{\scriptsize}
\tablewidth{18.0truecm}   
\tablecaption{Range of Predicted Fields\label{results2}}
\tablehead{
   \colhead{}&
   \multicolumn{2}{c}{VBJ}&
   \colhead{}&
   \multicolumn{2}{c}{HEG}&
   \colhead{}&
   \multicolumn{2}{c}{GHBC}&
   \colhead{}&
   \multicolumn{2}{c}{CG}&
   \colhead{}&
   \multicolumn{2}{c}{WG}\\[0.2ex]
   \cline{2-3}
   \cline{5-6}
   \cline{8-9}
   \cline{11-12}
   \cline{14-15}
   \colhead{}&
   \colhead{}&
   \colhead{}&
   \colhead{}&
   \colhead{}&
   \colhead{}&
   \colhead{}&
   \colhead{}&
   \colhead{}&
   \colhead{}&
   \colhead{}&
   \colhead{}&
   \colhead{}&
   \colhead{}&
   \colhead{}\\[0.2ex]
   \colhead{}&
   \colhead{$\dot{M} \times 10^8$}&
   \colhead{$B_{eq}$\tablenotemark{a}}&
   \colhead{}&
   \colhead{$\dot{M} \times 10^8$}&
   \colhead{$B_{eq}$\tablenotemark{a}}&
   \colhead{}&
   \colhead{$\dot{M} \times 10^8$}&
   \colhead{$B_{eq}$\tablenotemark{a}}&
   \colhead{}&
   \colhead{$\dot{M} \times 10^8$}&
   \colhead{$B_{eq}$\tablenotemark{a}}&
   \colhead{}&
   \colhead{$\dot{M} \times 10^8$}&
   \colhead{$B_{eq}$\tablenotemark{a}}\\[0.2ex]
   \colhead{Star}&
   \colhead{$(M_\odot {\rm yr}^{-1})$}&
   \colhead{(kG)}&
   \colhead{}&
   \colhead{$(M_\odot {\rm yr}^{-1})$}&
   \colhead{(kG)}&
   \colhead{}&
   \colhead{$(M_\odot {\rm yr}^{-1})$}&
   \colhead{(kG)}&
   \colhead{}&
   \colhead{$(M_\odot {\rm yr}^{-1})$}&
   \colhead{(kG)}&
   \colhead{}&
   \colhead{$(M_\odot {\rm yr}^{-1})$}&
   \colhead{(kG)}
}
\startdata
AA Tau  & 0.71 & 1.56 & & 12.6 & 4.04 & & 0.33 & 0.96 & & 0.40 & 1.05 & & 0.65 & 2.34 \\
BP Tau  & 2.43 & 2.46 & & 15.8 & 4.06 & & 2.88 & 1.62 & & 2.30 & 1.45 & & 1.32 & 1.81 \\
CY Tau  &\nodata&\nodata& & 0.63 & 2.62 & & 0.75 & 1.38 & & 0.80 & 1.43 & & 0.14 & 0.75 \\
DE Tau  & 18.0 & 0.07 & & 31.6 & 1.18 & & 2.64 & 0.49 & & 4.60 & 0.65 & & 4.07 & 0.62 \\
DF Tau  & 21.9 & 0.37 & &125.9 & 0.67 & & 17.7 & 0.57 & & 20.8 & 0.62 & & 1.00 & 2.03 \\
DG Tau  &\nodata&\nodata& &199.5 & 9.09 & &\nodata&\nodata& & 91.9 & 7.36 & & 4.57 & 2.44 \\
DH Tau  & 2.83 & 0.25 & &\nodata&\nodata& &\nodata&\nodata& &\nodata&\nodata& & 0.11 & 0.93 \\
DK Tau  & 0.61 & 1.93 & & 39.8 & 2.19 & & 3.79 & 0.95 & & 0.50 & 0.35 & &\nodata&\nodata\\
DN Tau  & 0.13 & 0.15 & & 3.20 & 0.84 & & 0.35 & 0.30 & & 0.20 & 0.23 & & 0.19 & 0.38 \\
GG Tau A& 3.01 & 1.48 & & 20.0 & 1.41 & & 1.75 & 1.05 & & 2.50 & 1.26 & & 1.26 & 1.79 \\
GI Tau  & 2.02 & 9.16 & & 12.6 & 1.06 & & 0.96 & 1.70 & & 0.60 & 1.35 & & 0.83 & 1.19 \\
GK Tau  & 0.05 & 0.23 & & 6.30 & 0.86 & & 0.64 & 0.32 & & 1.40 & 0.48 & & 0.65 & 0.57 \\
GM Aur  & 0.74 & 4.30 & & 2.50 & 5.20 & & 0.96 & 2.34 & & 0.70 & 2.00 & & 0.66 & 6.86 \\
 T Tau  &\nodata&\nodata& &\nodata&\nodata& &\nodata&\nodata& &\nodata&\nodata& & 3.16 & 0.39 \\
{\it Cor}\tablenotemark{b}& 0.27& 0.43 & & 0.21 & 0.51 & & 0.07 & 0.85 & & 0.21 & 0.52 & & 0.18 & 0.55 \\
\enddata
\tablenotetext{a}{The predicted field strength is the equatorial field
strength of an assumed dipole field using the Shu et al. (1994) model.}
\tablenotetext{b}{Given on this line in each $\dot M$ column is the value
of the correlation coefficient, $r$, between the predicted field strenghts 
and the observed mean field strengths.  Given on this line in each $B_{eq}$
column is the associated false alarm probability.}
\end{deluxetable}

\clearpage

\begin{deluxetable}{lcccc}
\tablewidth{6.0truecm}   
\tablecaption{Correlations of Derived Magnetic Parameters with Stellar Parameters\label{stelpar}}
\tablehead{
   \colhead{Quantities}&
   \colhead{}&
   \colhead{}\\[0.2ex]
   \colhead{Compared}&
   \colhead{$r$}&
   \colhead{$f_p$}
}
\startdata
$\bar B$ vs $P_{rot}^{-1}$ & 0.22 & 0.44 \\
$\Phi $ vs $P_{rot}^{-1}$ & 0.05 & 0.85 \\
$\bar B$ vs $\tau_c P_{rot}^{-1}$ & 0.23 & 0.41 \\
$\Phi $ vs $\tau_c P_{rot}^{-1}$ & $-0.24$ & 0.40 \\
$\bar B$ vs $\tau_c$ & 0.03 & 0.92 \\
$\Phi $ vs $\tau_c$ & $-0.57$ & 0.03 \\
$\bar B$ vs $T_{eff}$ & 0.26 & 0.36 \\
$\Phi $ vs $T_{eff}$ & 0.37 & 0.18 \\
$\bar B$ vs $M_*$ & 0.42 & 0.12 \\
$\Phi $ vs $M_*$ & 0.20 & 0.48 \\
$\bar B$ vs $L_*$ & 0.10 & 0.73 \\
$\Phi $ vs $L_*$ & 0.66 & $<0.01$ \\
$\bar B$ vs Age & 0.23 & 0.41 \\
$\Phi $ vs Age & $-0.31$ & 0.26 \\
\enddata
\tablecomments{Convective turnover times, $\tau_c$, were kindly computed 
by Y.-C. Kim, based on models described in Kim \& Demarque (1996) and 
Preibisch et al. (2005).}
\end{deluxetable}

\clearpage

\clearpage

\clearpage

\end{document}